\documentclass[prl,twocolumn,showpacs]{revtex4}
\newcommand{\be}{\begin{equation}}
\newcommand{\ee}{\end{equation}}
\newcommand{\bea}{\begin{eqnarray}}
\newcommand{\eea}{\end{eqnarray}}
\usepackage{graphicx}
\usepackage{epsfig}
\usepackage{bm}

\begin{document}
\title{Reconstruction of interacting dark energy models from parameterizations}
\author{R. Rosenfeld}
\affiliation{Instituto de F\'\i sica Te\'orica - UNESP \\ Rua
Pamplona, 145, 01405-900, S\~{a}o Paulo, SP, Brazil}

\begin{abstract}
Models with interacting dark energy can alleviate the cosmic coincidence problem
by allowing dark matter and dark energy to evolve in a similar fashion.
At a fundamental level, these models are specified by choosing a functional form
for the scalar potential and for the interaction term. However, in order to compare
to observational data it is usually more convenient to use parameterizations of the dark energy equation of
state and the evolution of the dark matter energy density. Once the relevant parameters are fitted
it is important to obtain the shape of the fundamental functions.
In this paper I show how to reconstruct the scalar potential and
the scalar interaction with dark matter from general
parameterizations. I give a few examples and show that it is possible for the effective
equation of state for the scalar field to cross the phantom
barrier when interactions are allowed. I analyze the uncertainties in the reconstructed potential 
arising from foreseen errors in the estimation of fit parameters and point out that a Yukawa-like
linear interaction results from a simple parameterization of the coupling.
\end{abstract}
 \preprint{IFT-P.xx/2006}

\maketitle

\section{Introduction}

Recent data from type Ia supernovae (SNIa) \cite{Riess,SNLS}, cosmic
microwave background (CMB) \cite{WMAP3y} and large scale structure (LSS)
\cite{SDSS} all point to the fact that the universe has recently
entered into a stage of accelerated expansion. This is an
unexpected and revolutionary discovery that calls for new physics,
since gravity is always an attractive force.

The simplest possibility to explain the acceleration of the
universe is to postulate the existence of a cosmological constant
of the right magnitude. While this assumption is still compatible
with all available data, it is unsatisfactory on the grounds of
the huge amount of fine tuning required. Hence, exploratory models
with scalar fields possessing a varying non-zero vacuum energy density,
usually called quintessence models \cite{reviews}, have been
studied as an alternative to the cosmological constant solution.
One of the most important tasks ahead of observational cosmology
is to devise methods and gather data to successfully and
unequivocally distinguish between these two possibilities. Finding
evidence for an evolving vacuum energy would be one of the
greatest discoveries of the century.

What is the appropriate scalar potential that could reproduce the
observed data? There has been a large number of papers dealing
with the possibility of reconstructing the potential directly from
the data
\cite{NakamuraChiba,HutererTurner,Starobinsky,SainiEtAl,MBS,Shinji,HutererPeiris,SLP,LiHolzCooray} and
it is fair to say that any statement about the form of the
potential depends upon some level of parameterization.
The procedure has also been extended to reconstruct the potential and
the form of the scalar-gravity coupling in the Jordan frame of
a scalar-tensor gravity theory from data on luminosity distance and
linear density perturbation \cite{BoisseauEtAl}.
More recently Guo, Ohta and Zhang \cite{GOZ} studied the reconstruction of the scalar potential
directly from parameterizations of the equation of state.

An intriguing possibility is that the scalar field is not totally
decoupled from our world \cite{CarrolAmendola}. Although the coupling to
baryonic matter is severely constrained, it is still possible to
allow for a coupling to non-baryonic dark matter. From a
lagrangian point of view, these couplings could be of the form
$W(\phi) m_0 \bar{\psi} \psi$ or $W(\phi) m_0^2 \varphi^2$ for a
fermionic or bosonic dark matter represented by $\psi$ and
$\varphi$ respectively, where the function $W$ of the quintessence
field $\phi$ can in principle be arbitrary. In this scenario, the
mass of the dark matter particles evolves according to some
function of the dark energy field $\phi$, leading to an effective
equation-of-state for the dark matter.

I will consider exploratory models for dark energy with a
canonical scalar field coupled to dark matter. The lagrangian for
this class of models is specified by the choice of two functions
of the scalar field: the scalar field potential $V(\phi)$ and
$W(\phi)$, the function that characterizes the coupling to dark
matter. Specific forms for the scalar potential and for the
interaction, such as power-law or exponential functions, have been
extensively studied in the literature
\cite{vamps,DasCorasanitiKhoury,HueyWendt}.

However, instead of postulating a concrete model by choosing
definite parametric forms for $V(\phi)$ and $W(\phi)$, it is often
more convenient and completely equivalent to introduce a
time-dependent parameterization for the dark energy equation of
state $w_{DE}(a)$ and for a coupling function $\delta(a)$, where
$a(t)$ is the scale factor of the universe. In fact, this approach
is widely used in the uncoupled case in order to test for the time-variation of dark
energy since it is much easier to compare to SNIa observations
\cite{parametrizations}.
For the coupled case, an analysis of the impact of interactions on SNIa
observations was performed using a specific coupling either
in the Einstein \cite{AGP} or Jordan \cite{NP} frame of a scalar-tensor theory.
More recently, parameterizations of $w_{DE}(a)$ and $\delta(a)$ were directly used
to study the effects of DE-DM interactions on fits from SNIa data \cite{us}.

Given a parameterization for $w_{DE}(a)$ and $\delta(a)$ with
parameters fitted by observations, it is important to investigate
the shape of the potential and interaction functions that results
in those parameterizations. They define the fundamental model behind the parameterizations.
In this article I will show how to
reconstruct the scalar potential and the interaction from the
general parameterized forms of the dark energy equation of state
$w_{DE}(a)$ and a coupling function $\delta(a)$ and provide a few concrete
examples.

\section{Reconstruction}

The reconstruction program proposed here generalizes the one
first developed by Ellis and Madsen \cite{EllisMadsen}. I consider a
spatially flat universe composed of three perfect fluids, namely
dark energy, non-baryonic dark matter and baryons. The dark matter
and baryons are non-relativistic pressureless fluids and
Einstein's equations result in:
\begin{eqnarray}
\label{eq:einstein}
H^2 &=& \frac{8 \pi G}{3} (\rho_\phi + \rho_{DM} + \rho_b) \\ \nonumber
\dot{H} + H^2 &=& -\frac{4 \pi G}{3} (\rho_\phi + \rho_{DM} + \rho_b + 3 p_\phi)
\end{eqnarray}
where $H = \dot{a}/{a}$ and the dark energy is described by a scalar field $\phi$
with energy density and pressure given by:
\begin{equation}
\label{eq:rhophi}
\rho_\phi = \frac{1}{2} \dot{\phi}^2 + V(\phi) \;, \;\;\;
p_\phi =  \frac{1}{2} \dot{\phi}^2 - V(\phi)
\end{equation}
with a equation of state $w_\phi$ defined by $w_\phi = p_\phi/\rho_\phi$.

Conservation of the stress-energy tensor requires that the total energy density
and pressure obey:
\begin{equation}
\label{eq:rhoTotal}
\dot{\rho}_T + 3 H (\rho_T + p_T) = 0.
\end{equation}
Introducing the coupling function $\delta(a)$ between dark energy and dark matter as
\begin{equation}
\label{eq:delta}
\delta(a) = \frac{d \ln m_\psi(a)}{d \ln a}
\end{equation}
results in the
following equation for the evolution of the DM energy density
$\rho_{DM}$ \cite{amendola}:
\begin{equation}
\dot{\rho}_{DM} + 3 H \rho_{DM} - \delta(a) H \rho_{DM} = 0.
\label{rhoDM}
\end{equation}
Also, conservation of baryon number requires:
\begin{equation}
\dot{\rho}_b + 3 H \rho_b = 0
\end{equation}
and  eq.~(\ref{eq:rhoTotal}) then implies that the dark energy density should obey
\begin{equation}
\dot{\rho}_{\phi} + 3 H (\rho_{\phi} + p_\phi ) + \delta(a) H \rho_{DM} = 0.
\label{rhoDE}
\end{equation}

Notice that the parameterization eq.~(\ref{eq:delta}) implies
\begin{equation}
\label{eq:Wreconstruction}
W(\phi(a)) = e^{-\int_{a}^{1} \delta(a') d \ln a' }
\end{equation}
normalized such that $W(\phi(a=1)) = 1$.

Combining
eqs.~(\ref{eq:rhophi},\ref{rhoDE},\ref{eq:Wreconstruction}) one
obtains a modified Klein-Gordon equation for the scalar field:
\begin{equation}
\ddot{\phi} + 3 H \dot{\phi} + \left( \frac{d V}{d \phi} +
\frac{\rho_{DM}^{(0)}}{a^3} \frac{d W}{d \phi} \right) = 0,
\end{equation}
in agreement with Das, Corasaniti and Khoury \cite{DasCorasanitiKhoury}.

One can now proceed to reconstruct the potential and the interaction for
a given parameterization of the equation of state $w(a)$ and the interaction
$\delta(a)$. The first step is to find the time variation of dark matter
energy density, which is easily obtained by solving eq.~(\ref{rhoDM}):
\begin{equation}
\rho_{DM}(a) = \rho_{DM}^{(0)} a^{-3} e^{-\int_{a}^{1} \delta(a') d \ln a' },
\end{equation}
where $\rho_{DM}^{(0)}$ is the non-baryonic DM energy density today.
It is more useful to work with the variable $u = \ln a$, and one can write
\begin{equation}
\rho_{DM}(u) = \rho_{DM}^{(0)} e^{-3 u}  e^{-\int_u^{0} \delta(u') d u' }.
\label{DMSol}
\end{equation}
The second step is to substitute $\rho_{DM}(u)$ into eq.~(\ref{rhoDE}), which in terms of $u$
reads:
\begin{equation}
\rho_\phi^\prime(u) + 3 (1 + w_\phi(u)) \rho_\phi(u) + \delta(u) \rho_{DM}(u) = 0,
\end{equation}
where $^\prime = d/du$, and find a solution $\rho_\phi(u)$ with initial condition
$\rho_\phi(u=0) =\rho_{\phi}^{(0)}$, with  $\rho_{\phi}^{(0)}$ being the dark energy density today.

In the third step one constructs the Hubble parameter:
\begin{equation}
\frac{H^2(u)}{H_0^2} = \Omega_b e^{-3 u} + \Omega_{DM} e^{-3 u}  e^{-\int_u^{0} \delta(u') d u' } +
\Omega_{\phi} f(u),
\end{equation}
where $\Omega_X = \rho_X^{(0)}/\rho_c^{(0)}$, the critical density today is
$\rho_c^{(0)} = 3 H_0^2/(8 \pi G)$ and $H_0$ is the Hubble constant. The function $f(u)$ that determines
the evolution of the dark energy density is in general obtained numerically.

Having obtained the Hubble parameter, the fourth step consists in
solving the evolution equation for the scalar field obtained from
eqs.~(\ref{eq:einstein},\ref{eq:rhophi}):
\begin{equation}
\left( \frac{d \tilde{\phi}}{d u} \right)^2 =
- \frac{1}{4 \pi} \left( \frac{d \ln H(u)}{d u} + \frac{3}{2} (\Omega_{DM}(u) + \Omega_b(u)) \right),
\label{eq:phi}
\end{equation}
where $\tilde{\phi} = \phi/M_{Pl}$ is the scalar field in units of the Planck
mass ($M_{Pl} = 1/\sqrt{G}$) and
\begin{equation}
\Omega_{DM,b}(u) = \frac{\rho_{DM,b}(u)}{\rho_\phi(u) + \rho_{DM}(u) + \rho_b(u)}.
\end{equation}
In the fifth step one numerically inverts the solution
$\tilde{\phi}(u)$ in order to determine $u(\tilde{\phi})$ to
finally obtain
\begin{eqnarray}
\tilde{V}(\tilde{\phi})\equiv
\frac{V(u(\tilde{\phi}))}{\rho_c^{(0)}} &=& \left( \frac{1}{3}
\frac{H(u)}{H_0} \frac{d H/H_0}{d u} + \frac{H^2(u)}{H^2_0} -
\right. \\ \nonumber && \left. \frac{1}{2} \Omega_b e^{-3u} -
\frac{1}{2} \Omega_{DM} e^{-3u} e^{-\int_u^{0} \delta(u') d u' }
\right)
\end{eqnarray}
and
\begin{equation}
\label{eq:Wreconstruction1}
W(u(\tilde{\phi})) = e^{-\int_u^{0} \delta(u') d u' }.
\end{equation}
This completes the reconstruction procedure. I will now work out some
examples of this procedure. I adopt $\Omega_\phi = 0.7$, $\Omega_{DM} = 0.25$ and
$\Omega_b = 0.05$ in the following. In all examples I integrate the field equation
starting from $u_{i}=-1.8$, corresponding to $z_i=5.05$ and I arbitrarily
set $\tilde{\phi}(u_i) = -1$.

\section{Examples}

\subsection{Constant $w_\phi$ and $\delta$}

\begin{figure}
\vspace{0.5cm}
\includegraphics[scale=0.6]{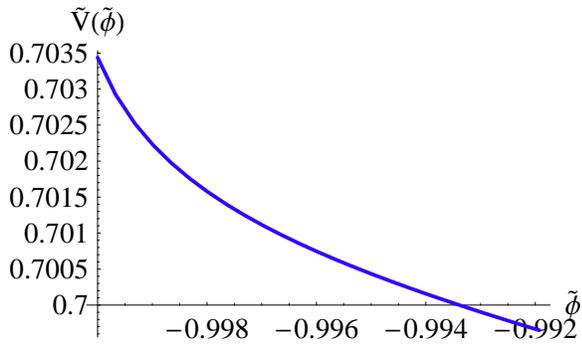}%
\caption{Reconstructed potential for $w_\phi = -0.99$ and
$\delta=0$. \label{fig:V099} }
\end{figure}

I start by considering the simple example of a constant equation
of state $w_\phi$ and constant coupling $\delta$. In order to gain
some intuition, I first show in fig.~\ref{fig:V099} the
reconstructed potential for $w_0 = -0.99$ and $\delta=0$. As
expected, the scalar field evolves slowly in a very flat
potential. Notice that the potential energy is slightly below $0.7
\rho_c$ today due to the small kinetic energy of the scalar field.

%
%%%%%%%%%%%%%%%%%%%%%%%%%%%%%%%%%%%%%%%%%%%%%%%%%%%%%%%%%%%%%%%%%%%%%%%%%%%%
%%%%%%%%%%%%%%%%%%%%%%%%%%%%%%%%%%%%%%%%%%%%%%%%%%%%%%%%%%%%%%%%%%%%%%%%%%%%%

I now turn on the interaction. In this case one
has \cite{AmendolaCamposRosenfeld}:
\begin{equation}
\rho_{DM}(a) = \rho_{DM}^{(0)} a^{-3+\delta},
\label{DMSol1}
\end{equation}
and the solution to eq.(\ref{rhoDE}) is:
\begin{equation}
\rho_{\phi}(a) = \rho_{\phi}^{(0)} a^{-3(1 + w_\phi)} +
\frac{\delta}{\delta + 3 w_\phi} \rho_{DM}^{(0)} \left(a^{-3(1 +
w_\phi)} - a^{-3+\delta} \right). \label{DESolConstantW}
\end{equation}
The first term of the solution is the usual evolution of DE
without the coupling to DM. From this solution it is easy to see
that one must require a positive value of the coupling $\delta>0$
in order to have a consistent positive value of $\rho_{\phi}$ for
earlier epochs of the universe. This feature remains in the case
of varying $w_\phi$ and in the rest of the paper I will assume
that $\delta$ is positive.

\begin{figure}
\vspace{0.5cm}
\includegraphics[scale=0.55]{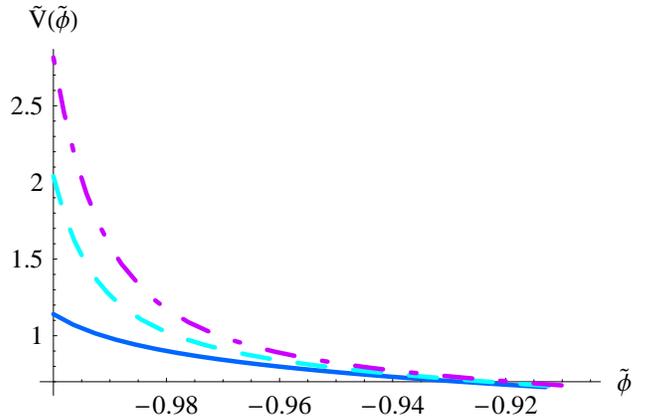}%
\caption{Reconstructed potential for $w_\phi = -0.9$ and
$\delta=0$ (solid), $\delta=0.05$ (dashed) and $\delta=0.1$ (dot-dashed). \label{fig:V09Delta} }
\end{figure}

In fig.~\ref{fig:V09Delta} I show the effects of the coupling in
the reconstructed potential. I use $w_0 = -0.9$ and $\delta=0,
0.05$ and $0.1$. The scalar potential becomes steeper for larger
values of the coupling due to the different dynamics introduced by
the DE-DM coupling.

%
%%%%%%%%%%%%%%%%%%%%%%%%%%%%%%%%%%%%%%%%%%%%%%%%%%%%%%%%%%%%%%%%%%%%%%%%%%%%
%%%%%%%%%%%%%%%%%%%%%%%%%%%%%%%%%%%%%%%%%%%%%%%%%%%%%%%%%%%%%%%%%%%%%%%%%%%%%
%

One might have expected that the introduction of the coupling
could allow the possibility of having a phantom equation of state,
$w_\phi <-1$. However, it is easy to show that one can write
eq.~(\ref{eq:phi}) for both the uncoupled and coupled cases as:
\begin{equation}
\left( \frac{d \tilde{\phi}}{d u} \right)^2 =
\frac{3}{8 \pi} \left(1+w_\phi(u) \right) \Omega_\phi(u)
\label{eq:phi1}
\end{equation}
in the general case of a time-varying equation of state and hence,
in order to have a real scalar field one must consider only
$w_\phi(u) \ge -1$.

\begin{figure}
\vspace{0.5cm}
\includegraphics[scale=0.55]{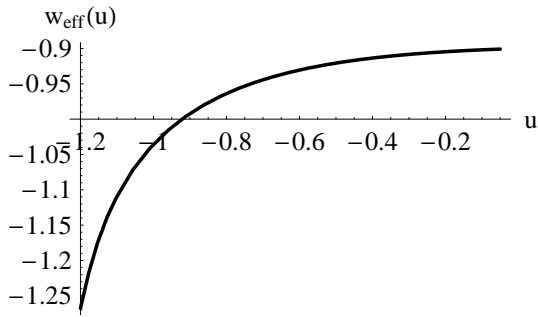}%
\caption{Effective equation of state for $w=-0.9$ and $\delta = 0.1$.
\label{fig:weff} }
\end{figure}

However, the fitter who is unaware of the interaction would instead find
an effective equation of state $w_{eff}(u)$ defined implicitly by:
\begin{equation}
\rho_\phi^{(0)} e^{-3 (1 + w_{eff}(u)) u} + \rho_{DM}^{(0)} e^{-3 u} =
\rho_\phi(u) + \rho_{DM}^{(0)} e^{(-3+\delta) u}.
\label{eq:weff}
\end{equation}
In fig.~\ref{fig:weff} I show that the effective equation of
state {\it can} in fact cross the phantom barrier. This possibility was
also recently pointed out in the context of scalar-tensor
theories of gravity \cite{tricked}.

%
%%%%%%%%%%%%%%%%%%%%%%%%%%%%%%%%%%%%%%%%%%%%%%%%%%%%%%%%%%%%%%%%%%%%%%%%%%%%
%%%%%%%%%%%%%%%%%%%%%%%%%%%%%%%%%%%%%%%%%%%%%%%%%%%%%%%%%%%%%%%%%%%%%%%%%%%%%
%

\begin{figure}
\vspace{0.5cm}
\includegraphics[scale=0.45]{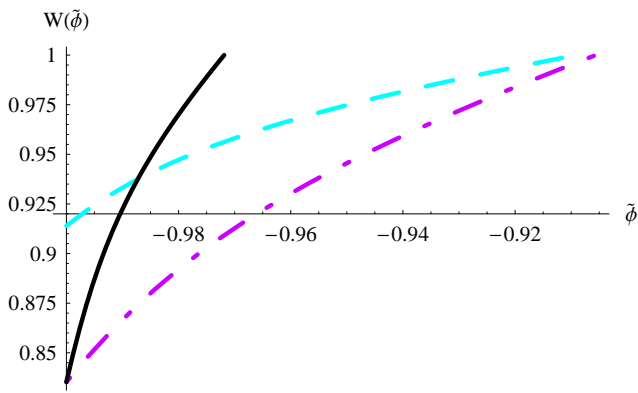}%
\caption{Reconstructed interaction for $w=-0.9$ and $\delta = 0.05$ (dashed line),
$\delta = 0.1$ (dot-dashed line) and $w=-0.99$ and $\delta=0.1$ (solid line).
\label{fig:W} }
\end{figure}

One can also easily reconstruct the interaction $W(\tilde{\phi})$ in this simple case:
\begin{equation}
W(\tilde{\phi}(u)) = e^{\delta u},
\end{equation}
which of course guarantees that $W(\tilde{\phi}(u=0))=1$ as
required. In fig.~\ref{fig:W} I plot the reconstructed
interaction term for $w=-0.9$ and $\delta = 0.05, \; 0.1$ and also
for $w=-0.99$ and $\delta=0.1$. Notice that the interaction increases with time,
corresponding to a mass that decreases with increasing redshift. One can
see that the interaction term becomes steeper with increasing
$\delta$ and decreasing $w$.

%
%%%%%%%%%%%%%%%%%%%%%%%%%%%%%%%%%%%%%%%%%%%%%%%%%%%%%%%%%%%%%%%%%%%%%%%%%%%%
%%%%%%%%%%%%%%%%%%%%%%%%%%%%%%%%%%%%%%%%%%%%%%%%%%%%%%%%%%%%%%%%%%%%%%%%%%%%%
%

\subsection{Variable $w_\phi$ and constant $\delta$}

I now analyze the potential reconstructed from the commonly used
2-parameter description of the equation of state \cite{paramet}:
\begin{equation}
w_\phi(a) = w_0 + w_1 (1-a).
\end{equation}
A recent fit to SNIa, CMB and LSS data obtained (without DE
perturbation) \cite{fit}:
\begin{equation}
w_0 = -1.098^{+ 0.078}_{- 0.080}, \;\;\; w_1 =
0.416^{+0.293}_{-0.153}
\end{equation}

\begin{figure}
\vspace{0.5cm}
\includegraphics[scale=0.45]{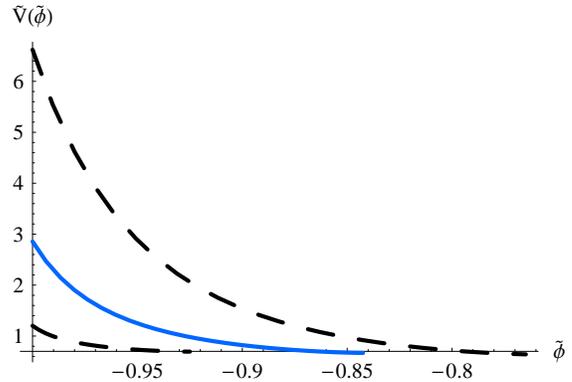}%
\caption{Reconstructed potential for $w_0=-0.91$, $w_1 = 0.4$ and $\delta = 0$ (solid line).
The allowed region from the $1-\sigma$ uncertainties in the estimation of the
equation of state parameters is the region between the dashed lines.
\label{fig:V_variable_noInt} }
\end{figure}

Hence I will assume that it is possible to fit the 2-parameter equation of state with a
1-$\sigma$ precision of
$\delta w_0 = 0.08$ and $\delta w_1 = 0.20$. For illustration purposes, I will take the central
values $w_0 = -0.91$ and $w_1 = 0.4$ and study the effects of dark energy interaction in
the reconstruction of the potential.
In fig.~\ref{fig:V_variable_noInt} I show the allowed region in the potential form arising
from the 1-$\sigma$ uncertainties in the estimation of the
equation of state parameters. This region therefore gives an idea of the uncertainty one can expect 
in the reconstructed potential given the foreseen errors in the fit parameters.

%
%%%%%%%%%%%%%%%%%%%%%%%%%%%%%%%%%%%%%%%%%%%%%%%%%%%%%%%%%%%%%%%%%%%%%%%%%%%%
%%%%%%%%%%%%%%%%%%%%%%%%%%%%%%%%%%%%%%%%%%%%%%%%%%%%%%%%%%%%%%%%%%%%%%%%%%%%%
%

\begin{figure}
\vspace{0.5cm}
\includegraphics[scale=0.45]{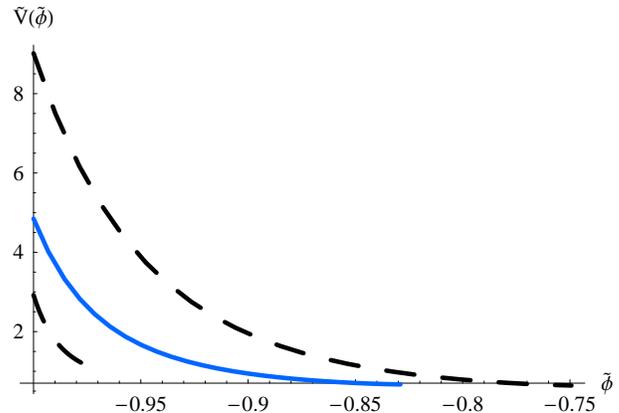}%
\caption{Reconstructed potential for $w_0=-0.91$, $w_1 = 0.4$ and
$\delta = 0.1$ (solid line). The allowed region from the
$1-\sigma$ uncertainties in the estimation of the equation of
state parameters is the region between the dashed lines.
\label{fig:V_variable_Int} }
\end{figure}

In fig.~\ref{fig:V_variable_Int} I show the effect of interaction
with $\delta = 0.1$ on the allowed region in the potential form
arising from the $1-\sigma$ uncertainties in the estimation of the
equation of state parameters.

%
%%%%%%%%%%%%%%%%%%%%%%%%%%%%%%%%%%%%%%%%%%%%%%%%%%%%%%%%%%%%%%%%%%%%%%%%%%%%
%%%%%%%%%%%%%%%%%%%%%%%%%%%%%%%%%%%%%%%%%%%%%%%%%%%%%%%%%%%%%%%%%%%%%%%%%%%%%
%

\subsection{Constant $w_\phi$ and variable $\delta$}

Finally, I analyze the case of a constant equation of state and an interaction
parameterized by the function \cite{Waga}:
\begin{equation}
\delta(a) = \delta_0 \frac{2 a}{1+a^2},
\end{equation}
which is well behaved in the past as well as in the future and $\delta(a=1) = \delta_0$.
In this case one finds:
\begin{equation}
\rho_{DM}(a) = \rho_{DM}^{(0)} a^{-3} e^{2 \delta_0 (\arctan(a) -
\pi/4)}
\end{equation}

\begin{figure}
\vspace{0.5cm}
\includegraphics[scale=0.45]{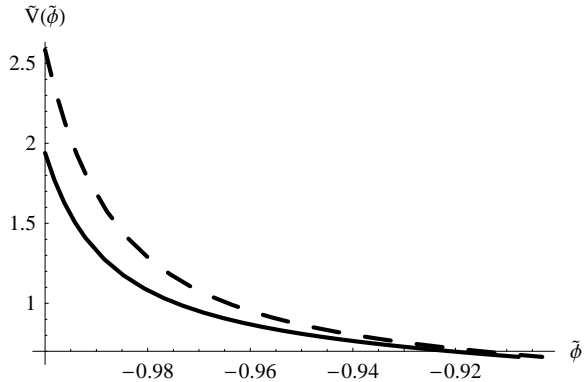}%
\caption{Reconstructed potential for constant $w=-0.9$, and $\delta_0 = 0.1$ (solid line)
and $\delta_0 = 0.2$ (dashed line).
\label{fig:V09VaryingDelta} }
\end{figure}

\begin{figure}
\vspace{0.5cm}
\includegraphics[scale=0.45]{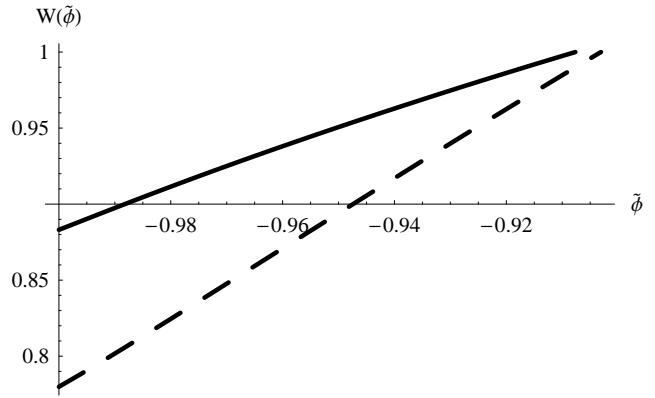}%
\caption{Reconstructed interaction function for constant $w=-0.9$, and $\delta_0 = 0.1$ (solid line)
and $\delta_0 = 0.2$ (dashed line).
\label{fig:W09VaryingDelta} }
\end{figure}

The reconstructed potential is shown in fig.~\ref{fig:V09VaryingDelta} for $w_\phi = -0.9$
and $\delta_0= 0.1, 0.2$.  
The interaction function is given by
\begin{equation}
W(u(\tilde{\phi})) = e^{2 \delta_0 (\arctan(e^{u(\tilde{\phi})}) -
\pi/4)}
\end{equation}
and it is plotted in fig.~\ref{fig:W09VaryingDelta}.
Notice that in this example the interaction function has an approximate 
Yukawa-like linear form, even more so for the $\delta_0= 0.2$ case.  

%
%%%%%%%%%%%%%%%%%%%%%%%%%%%%%%%%%%%%%%%%%%%%%%%%%%%%%%%%%%%%%%%%%%%%%%%%%%%%
%%%%%%%%%%%%%%%%%%%%%%%%%%%%%%%%%%%%%%%%%%%%%%%%%%%%%%%%%%%%%%%%%%%%%%%%%%%%%
%

%
%%%%%%%%%%%%%%%%%%%%%%%%%%%%%%%%%%%%%%%%%%%%%%%%%%%%%%%%%%%%%%%%%%%%%%%%%%%%
%%%%%%%%%%%%%%%%%%%%%%%%%%%%%%%%%%%%%%%%%%%%%%%%%%%%%%%%%%%%%%%%%%%%%%%%%%%%%
%

\section{Conclusions}

It is natural to study models where the dark energy field is not totally 
separated from the rest of the world. In principle, it can interact with dark matter and this
possibility has interesting consequences in the evolution of the universe, such as a mass-varying
cold dark matter particle with a non-zero effective equation of state.

In order to compare these models to observations it is more convenient to use 
parameterizations of the dark energy equation of
state and the coupling to dark matter. However, once the parameters are estimated it is important 
to find the fundamental lagrangian of the theory, that is, to determine the functional form 
of the scalar potential and its interaction with dark matter.

I showed in this paper how to reconstruct the scalar potential and
the scalar interaction with dark matter from general
parameterizations. For illustration purposes some examples are worked out.
Uncertainties in the reconstruction due to uncertainties in the estimation of parameters
are analyzed.
It is pointed out that the phantom barrier can be crossed if the fit does not
take into account interactions and that  a Yukawa-like
linear interaction results from a simple parameterization of the coupling.

More precise data from CMB, SNIa and LSS from different collaborations 
is expected to arrive in the near future. These data can be used to estimate 
the parameters of simple parameterizations and the procedure shown here can at least
point towards the underlying fundamental model describing our universe.

\section*{Acknowledgments}
I would like to thank G. C. Campos for keeping asking the question that led to this
investigation, and L. R. Abramo and U. Fran\c{c}a for a careful reading of the manuscript.
This work was partially supported by a CNPq research grant 305055/2003-8.


\begin{thebibliography}{11}
\bibitem{Waga}
I would like to thank Ioav Waga for suggesting this parameterization for
the coupling $\delta(a)$.
\end{thebibliography}
\end{document}